# Statistical analysis of the effect of the current, potential and proposed rules of a game in tennis

**G. Szigeti**

*Szent István University, Gödöllő, Hungary EU,*
*szigeti.gabor@gtk.szie.hu*

*Abstract*

The tennis match and its scoring units, point, game and set have been already analyzed as a Markov chain in different aspects. Some selected results from this rich literature will be mentioned in the paper. We examine a new position, our major problem involves the position of a spectator: in top class tennis, especially in the men's game the predictability of the core unit, the game is too high.

With the aid of mathematical modelling (basic tool is the random walk with absorbing barriers) we derive subsequent formulas to study the effect of different versions of possible rules. For different rules the probability of winning a game, the probability of break point occurrence, the mathematical expectation of the number of rallies (points) and, the mathematical expectation of the number of break points in a game are expressed. We check these rules against ATP statistics for the Top-200 men players.

In conclusion, we suggest a slight but essential modification for the rule of a tennis game, namely , second service ( in case of a first service fault) is to be allowed only at the first three points (rallies). This would partially preserve the traditions (server has an advantage in the modern game) and at the same time it would reduce the predictability of the game, significantly increasing in this way the excitement for the spectators.

Taking the suggestion seriously, using the proposed new rule of this paper, experimental tournaments are also recommended.

*Introduction*

Sports statistics improves every day and one can expect only positive results from this process, because it brings the statistical science close to a large new audience, teaching this way an unbiased, fact-based approach of thinking.

A lot of technical innovation is involved in data acquisition for different branches of sports, still, from an analytical point of view, sports statistics matures into its most developed form only, if it is not just a simple counting or aggregating of different elements in sport, but if using these data one can analyze certain aspects of the technical-tactical-strategical triad of a sport competition.

Sport statistics has been applied for many different purposes, such as improving betting performance, producing better results for individual players and for teams, finding the weaknesses of the opponent, studying the potential consequences of different modifications to the rules, avoiding injury etc..

In tennis, probably the earliest application of Markov chain belongs to R.W.Schutz (1970). T.J.Barnett (2002) improves it by calculating conditional probabilities, i.e. determining the chances of winning for a player, starting from a given score line. This can be applied for online betting purposes.

Particularly, the service action is also subject to study by sport-statistical literature. Using and analyzing a large data set, T.Barnett (2008) gives advices to players, how to utilize their capabilities. P. Norton (2002) shows, that the popular statement of "serving with the advantage of new balls" cannot be scientifically justified. Also he compares different Grand Slam surfaces from a service point of view. A set of new rules proposed by ITF (International Tennis Federation) is studied by G. Pollard (2002), with respect to the length of the match. In his paper we find the valuable for us remark ".. an increase in the percentage of service breaks may not be a bad thing for the game".

Here we come up with this kind of special problem in respect of tennis. It's generally accepted that the crowd's attention and excitement during longer periods of a tennis match is low. The reason is a too high predictability of --not the outcome of the total match-- but the outcome of the game within the match. It is not infrequent that the set progresses into six all without a single break game, followed by a decisive tie-break.

In modern tennis, especially in the men's game, with the improvement of the tennis rackets and with the increasing height of players the server has a serious advantage from the beginning of the rally. Increased body height could be a positive factor on the receiver's side as well, but the good dynamic properties of the up to date racket does not help to receive, because there is not enough time to produce a technically clean return shot.

Probably one reason for the overwhelming popularity of football (soccer) is that the weaker team also has a positive, i.e. not negligible, chance to win a game, as good scoring positions, worked out by a team, occur only relatively few times during the 90 minutes. In tennis, the high number of rallies make this event less likely, only a very few times, and only in very even contests does an experienced observer feel that the weaker player has won the match with the aid of Fortuna. Of course there is no need to try to change it, this should be perceived as the very nature of tennis . But let the stronger player win with a less predictable string of score line, reducing the number of games won by the server.

For this purpose we introduce different 'imaginary' rules and study their effect on the game with the aid of mathematical modelling. For our purpose, it is enough to analyze the game. The fact that without the jus-situation the game goes up to four points won by the same player, rather than the set, which goes up to six games, gives us a chance to find polynomial formulas, rather than the more common in the literature algorithmical solution. As a result, now, these polynomial formulas are easier to check and analyze than the alghoritmical solution.   Together with the probability of winning a game, and the mathematical expectation of the number of rallies, which are known e.g. from (Newton at al. 2009) and (Madurska 2012), we constructed formulas for the probability of  break point occurrence as well as for the mathematical expectation of the number of break points within a game. Because the newly proposed rules require new formulas, we gave the mathematical thinking (derivation), which leads to them.

Models with one variable are analyzed for the full range of initial probability of winning a point, while models of two independent variables, for the sake of conciseness,  are analyzed basically only for 'real world' data,  only for the ATP Top-200 data range.

***Real and potential rules of a tennis game***

A real tennis game, i.e. a game under the current rules continues  until one of the players  wins four points but with a positive difference of at least two points. We do not intend to change this. We introduce a number of possible tennis games, represented in  Figure 1, where the order of serves is subject to change (with the exception of the last row).

The player who serves first is called F. The other player's name is S, like first and second.  The first three rows in the figure  ( deuce games) are necessary only for the study of the 'deuce' situation, i.e. the situation  that after 6 points  the game progresses into a 3 to 3, deuce state. A-type deuce describes the real tennis; F serves all the time. Bj(1) and Bj(2) are two possible different modifications: both players serve in the order  indicated by the Figure 1. In a real game, the two Bj systems might not be identical, but in the mathematical model, because the outcome of rallies is assumed to be independent events ( in the sense of theory of probability), and  because deuce-type games might be over only in even number of points played, Bj(1) and Bj(2) give exactly the same model answer. We remark that the break point loses its concept and importance here, and  we shall not calculate it.

The arrows, both pointing down and up,  in Figure 1, indicate all the possible end points of a game. The upper line with certain combination of letters  F and S indicates the shortest repeating unit.

The T-type in the fourth row is the real, existing tennis game. B(1) and B(2) type tennis games are equivalent in the same sense as it was explained at Bj(1) and Bj(2) deuce-type, but now, this is true in a less self-evident way; the game might end either with an even number of points or in five points (with the score of 4:1) and in the latter case F gave three services in both versions (that would not be the case at 4:3).

C-type game reflects our major idea. Here we wish to balance the server's advantage  in a T-type game,  to some extent.  Here and only here ( in the seventh row) S does not mean that player S serves, but the idea is, that after a certain number of points played, F should not be allowed to give a second serve (in case of a first service fault), and as a consequence, now F has a lower chance to win the point. Again from a mathematical  model point of view, this is just a reduced (just a different)  chance to win the point (as if the other player would serve). The shape of this C-type, in which F has exactly three positive opportunities (sequence F,F,F,S,S,S,S,S,...) will be explained later. Three letter F will be introduced not arbitrarily, but as a result of model estimation based on real ATP data.

For the sake of completeness, we mention a further possibility, that player F has a limited, x number of points ( say x= 2,3 4..), occasions to commit a first service fault (i.e. give a second one in case of a fault). This would create a situation similar to the umpire challenging system from the player's side. Mathematical study of the effect of such a rule is fairly possible, but we do not recommend this, because it would be practically less feasible ( as the " player has two challenges remaining" statement from the chair umpire is somewhat awkward at ATP and WTA tennis).

***Model answer for different games***

All the different games will be considered as a Markov process. Rallies are independent, identically distributed, and basic probabilities are given as

P(player F wins a rally when he serves) = $p_F$ ( $p_F$ also marked by $p$ in case only F serves during the game).

P(player F wins a rally when S serves) = $p_S$ . Please note that while S is serving , the probability is still considered from the position of player F. We put

$$q = 1 - p, \quad q_F = 1 - p_F, \quad q_S = 1 - p_S .$$

Notations are collected at the end of the paper.

**Deuce-type section**

These cases (first three rows of Figure 1) are necessary to resolve the deuce situation, and their separate calculation is justified, because the results will be used during calculation of the more complete cases ( rows from 4 to 7 of Fig. 1).

At ***A-type model***, the game has 5 different states, as on (a) part of Figure 2, and, before the first rally, the 0 state (0:0, love all state) is a certain event. In each consecutive step (point) the state moves to the left by probability $p$ or to the right by probability $q = 1 - p$ . In the third row either the game is over (with $p^2$ probability in favor of F and $q^2$ of S) or it goes back to the 0 state, so for the total probability of player F winning the game, using the sum for a geometrical series, one can write

$$P_A(p) = p^2 + 2pq\{p^2 + 2pq[p^2 + 2pq(p^2 + 2pq(........))]\} =$$

$$= p^2[1 + 2pq + (2pq)^2 + (2pq)^3 + ...] = \frac{p^2}{1-2pq} = \frac{p^2}{1-2p+p^2+p^2}$$

And finally,

$$P_A(p) = \frac{p^2}{p^2+q^2} \tag{1}$$

The chance for S to win the game is given when $p$ substituted by $q$ (and $q$ by $p$ ). By this we have $P_A(p)+P_A(q)=1$. This shows that while the draw (infinite returning of the 0 state) is a logical possibility, its probability measure is zero.

When calculating the probability of the event, that a break point does occur (at least once) during an A-type game, we create the disjoint partition of the event, namely $\{\textit{breakpont occurs the first time as a result the } s^{th} \textit{ rally}\}, s = 1,3,5...$ , and, accordingly we have to cut the scheme of Figure 2 , part (a) at the state of +1 (i.e. at advantage of receiver) obtaining in this way part (b); so one can write

$$P_A^{br}(p) = q + pq\{q + pq[q + pq(q + pq(........))]\} =$$

$$= q[1 + pq + (pq)^2 + (pq)^3 + ...] = \frac{q}{1-pq} = \frac{q}{1-p+p^2} = \frac{q}{q+p^2} \tag{2}$$

At calculating the mathematical expectation of the number of points played, one considers that the game might be over only in 2, 4, 6... points. Introducing temporary notations $s =: 2pq$ and $Q =: p^2 + q^2$ ( by the sign =: we mean equation defining a new letter in order to shorten the derivation rather than a substantial equation) , and using the knowledge of the multiplication of a geometrical series by itself in the form

$$1 + 2r + 3\cdot r^2 + .... + n\cdot r^{n-1} + ... = \frac{1}{(1-r)^2} \quad \textit{if } |r| < 1$$

one can write

$$E_A(p) = 2Q + s\{4Q + s[6Q + s(8Q.....)]\} = 2Q[1 + 2s + 3s^2 + 4s^3 + ...] =$$

$$= \frac{2Q}{(1-s)^2} = \frac{2\cdot(p^2+q^2)}{(1-2pq)^2} = \frac{2\cdot(p^2+q^2)}{(p^2+q^2)^2} = \frac{2}{p^2+q^2} \quad (3)$$

When calculating the mathematical expectation of the number of break points during the A-type game, we must add up the probabilities of all the break points, because either the break situation directly continues with the final state (-2) or with a certain (reduced) probability it leads to a new break point:

$$E_A^{br}(p) = q + 2pq\{q + 2pq[q + 2pq(q + 2pq(........)]\} =$$

$$= q[1 + 2pq + (2pq)^2 + (2pq)^3 + ...] = \frac{q}{1-2pq} = \frac{q}{p^2+q^2} \quad (4)$$

***Bj-type deuce model*** is simpler to calculate for the Bj(1)-type. Here the game returns in two points to the deuce situation with a probability of $R =: p_F q_S + q_F p_S$ and F wins in two points with a probability of $p_F p_S$, so the total chance for winning the game by F can be written as

$$P_{Bj}(p_F, p_S) = p_S p_F + R\{p_S p_F + R[p_S p_F + R(p_S p_F + R(........)]\} =$$

$$= p_S p_F [1 + R + R^2 + R^3 + ...] = \frac{p_S p_F}{1 - p_F q_S - q_F p_S} = \frac{p_F p_S}{p_F p_S + q_F q_S}$$

$$P_{Bj}(p_F, p_S) = \frac{p_F p_S}{p_F p_S + q_F q_S} \quad (5)$$

,which is evidently a generalization of (1). Similarly to the derivation of (3), for Bj-type mathematical expectation we got

$$E_{Bj}(p_F, p_S) == \frac{2}{p_F p_S + q_F q_S} \quad (6)$$

Formula (5) has several properties worth mentioning

$$P_{Bj}(p_F, p_S) = P_{Bj}(p_S, p_F) \tag{7a}$$

$$P_{Bj}(\frac{1}{2}, p_S) = p_S \tag{7b}$$

$$P_{Bj}(p_F, p_S) + P_{Bj}(1-p_F, 1-p_S) = 1 \tag{7c}$$

$$P_{Bj}(p, 1-p) = \frac{1}{2} \tag{7d}$$

(7a) expresses the indifference of the order of serves for this type of game; (7b) easily checked by (6) but its meaning is less evident. It can be interpreted as fifty-fifty points are good only to add a point to the better player's score. (7c) describes , that one of the players' win is a certain event; (7d) is the situation, where both players have the same chance of winning the point and so winning the game. A further property is that (5) has a singularity at point $(p_F, p_S) = (1, 0)$ (also at $(p_F, p_S) = (0, 1)$ ) , the limit value is a function of the direction, with which $(p_F, p_S)$ 'curve' tends to this point. Limit does not exist otherwise.

### Complete games

The fact that **real tennis , T-type** and hypothetical B and C types are played at least up to four points does not represent any further mathematical difficulties other than rigorous calculations. Figure 3 shows the Markov process, and for example when calculating the chances for player F to win the game, one has only to sum up the individual favorable end states and the already obtained deuce formula (1) multiplied by the probability of the occurrence of this state

$$P_T(p) = p^4 + 4p^4q + 10p^4q^2 + 20p^3q^3 \cdot \frac{p^2}{p^2+q^2} \tag{8}$$

Equation (8) has a shape completely different from that of O'Malley (2008), but bringing them to a common canonical form, we proved that they represent exactly the same function of $p$ . His formulae is

$$P_T(p) = p^4 \cdot \left(15 - 4p - \frac{10p^2}{1-2p(1-p)}\right)$$

In our view, (8) has the advantage, that it directly reflects the game (F may win 4:0 with the probability of the first term, 4:1 with that of second ..etc.).

In a similar fashion, combining the ideas above , we get

$$P_T^{br}(p) = q^3 + 3pq^3 + 6p^2q^3 + 10p^3q^3 \cdot \frac{q}{q+p^2} \tag{9}$$

$$E_T(p) = 4(p^4+q^4) + 20pq(p^3+q^3) + 60p^2q^2(p^2+q^2) + \\ + 120p^3q^3 + 20p^3q^3 \cdot \frac{2}{p^2+q^2} \tag{10}$$

$$E_T^{br}(p) = q^3 + 4pq^3 + 10p^2q^3 + 20p^3q^3 \cdot \frac{q}{p^2+q^2} \tag{11}$$

For the ***B-type game*** the break point concept loses its original meaning, so we calculate only the formula for the main probability (F wins the game), and the number of rallies.

$$P_B(p_F, p_S) = p_F^2 p_S^2 + 2p_F^3 p_S q_S + 2p_F^2 p_S^2 q_F + \\ + 6p_F^2 p_S^2 q_F q_S + 3p_F^1 p_S^3 q_F^2 + p_F^3 p_S q_S^2 + \\ + \{9p_F^2 p_S q_F q_S^2 + 9p_F p_S^2 q_F^2 q_S + p_F^3 q_S^3 + p_S^3 q_F^3\} \cdot \frac{p_F p_S}{p_F p_S + q_F q_S} \tag{12}$$

Since we get results for the B and C type models in the form of ratio of polynomials, which formally contains four variables, but only two of them are independent, so for the sake of compactness, we introduce ‘mini-algebra’ in the form of

$$\mathrm{n} \cdot [\alpha, \beta, \gamma, \delta] =: n \cdot \mathrm{p}_S^{\alpha} \cdot q_S^{\beta} \cdot p_F^{\gamma} \cdot q_F^{\delta} \tag{13}$$

Here $\alpha + \gamma$ reflects the points of F, while $\beta + \delta$ reflects that of S. And, because we frequently have symmetrical terms, let us introduce a notation for this symmetricity operation

$$n\cdot\overline{\overline{[\alpha,\beta,\gamma,\delta]}} =: \quad \mathrm{n}\cdot[\alpha,\beta,\gamma,\delta]+\mathrm{n}\cdot[\beta,\alpha,\delta,\gamma]$$

(14)

Now, we may express (12) in this new form

$$P_B(p_F,p_S)=[2,0,2,0]+2[1,1,3,0]+2[2,0,2,1]+6[2,1,2,1]+ \\ +3[3,0,1,2]+[1,2,3,0]+\left\{9\cdot\overline{\overline{[1,2,2,1]}}+\overline{\overline{[3,0,0,3]}}\right\}\cdot\frac{[1,0,1,0]}{\overline{\overline{[1,0,1,0]}}}$$

(15)

In this style, formula for the mathematical expectation

$$E_B(p_F,p_S)=4\cdot\overline{\overline{[2,0,2,0]}}+5\cdot\left(2\overline{\overline{[2,0,2,1]}}+2\overline{\overline{[1,1,3,0]}}\right)+ \\ +6\cdot\left(3\overline{\overline{[3,0,1,2]}}+6\overline{\overline{[2,1,2,1]}}+\overline{\overline{[1,2,3,0]}}\right)+ \\ +\left\{9\cdot\overline{\overline{[1,2,2,1]}}+\overline{\overline{[3,0,0,3]}}\right\}\cdot\left(\frac{2}{\overline{\overline{[1,0,1,0]}}}+6\right)$$

(16)

Formula (16) reflects, that in four points F or S can only win 4:0, in 5 only 4:1 $(\alpha+\gamma:\beta+\delta)$ etc..

For the ***C-type game*** ( which will be justified later on) we calculate all the four quantities, because this model represents our main conclusion. We arrived at

$$P_C(p_F,p_S)=[1,0,3,0]+[1,1,3,0]+3[2,0,2,1]+[1,2,3,0]+ \\ +6[2,1,2,1]+3[3,0,1,2]+\left\{9\cdot\overline{\overline{[1,2,2,1]}}+\overline{\overline{[3,0,0,3]}}\right\}\cdot\frac{[2,0,0,0]}{\overline{\overline{[2,0,0,0]}}}$$

(17)

$$P_C^{br}(p_F,p_S)=[0,0,0,3]+3[0,1,1,2]+3[0,2,2,1]+3[1,1,1,2]+ \\ +\left\{6[1,2,2,1]+3[2,1,1,2]+[0,3,3,0]\right\}\cdot\frac{[0,1,0,0]}{[0,1,0,0]+[2,0,0,0]}$$

(18)

$$E_C(p_F, p_S) = 4 \cdot \overline{\overline{\overline{[1,0,3,0]}}} + 5 \cdot \left( \overline{\overline{\overline{[1,1,3,0]}}} + 3\overline{\overline{\overline{[2,0,2,1]}}} \right) + \\ + 6 \cdot \left( 3\overline{\overline{\overline{[3,0,1,2]}}} + 6\overline{\overline{\overline{[2,1,2,1]}}} + \overline{\overline{\overline{[1,2,3,0]}}} \right) + \\ + \left\{ 9 \cdot \overline{\overline{\overline{[1,2,2,1]}}} + \overline{\overline{\overline{[3,0,0,3]}}} \right\} \cdot \left( \frac{2}{\overline{\overline{\overline{[2,0,0,0]}}}} + 6 \right)$$

(19)

$$E_C^{br}(p_F, p_S) = [0,0,0,3] + [1,0,0,3] + 3[0,1,1,2] + [2,0,0,3] + 6[1,1,1,2] + \\ 3[0,2,2,1] + \left\{ 9 \cdot \overline{\overline{\overline{[1,2,2,1]}}} + \overline{\overline{\overline{[0,3,3,0]}}} \right\} \cdot \frac{[0,1,0,0]}{\overline{\overline{\overline{[2,0,0,0]}}}}$$

(20)

***Analysis of the models***

On the one hand, the models belonging to different rules are compared to each other, on the other hand, the models are checked against ATP Top-200 ( top 200 players of the period Jan.1991-Dec. 2016) statistics.

**Comparison of different rules**

Let us first see the consequence of the ‘at least four points’ addition to the ‘with two difference’ rule, i.e. let us compare A and T-type games.

In Figure 4 the graphs of formulas (1), (2), (8) and (9) are presented. Let us interpret the curves by tennis concepts. For the practical range $p \geq ½$ T-rule further reduces the chances for player S. In harmony with this, for the range $p \geq .4$ a break point becomes less likely, because after a lucky win of the first point by S, what follows is not a break point yet in T game. This case is different for $p < .4$, when S has good chances, then with the four-points-rule S is more likely to avoid the unlucky situation, when F wins too quickly, rather than S would win necessarily through a break point.

In Figure 5 the graphs of formulas (3), (4), (10) and (11) are presented. The highest mathematical expectation, i.e. the expected longest game duration is of course at $p = ½$ for T-type game, and that is 6¾. For A-type game, a flat maximum for the number of break points is at the value $p = \frac{2-\sqrt{2}}{2} = 0.293$. We mention here, that the $E_A(½) = 4$ relation is only a special case of a much more general, far from self-evident relation: in a deuce-type game but with n points of difference (random walk with absorbing barriers) , this mathematical expectation is $E_{..n..}(½) = n^2$, see in (Feller 1971) and (Szigeti at al. 2017).

For the $Bj$ and $B$-type model we have two independent variables. We found that for a relatively large set of parameters $(p_F, p_S)$, the game-winning probability of F is a strong function (i.e. nearly constant function) of the not too large difference in chances for point-winning of the two players. The order of service is indifferent, so it is enough to study the case, when $p_F \geq q_S$; $q_S = p_F - \delta$, $\delta \geq 0$ and $p_S = 1 - p_F + \delta$. Figure 6 contains the subsequent curves for the values $\delta = 0.;\ 0.05;\ 0.1;\ 0.2$. The pairs of curves $(P_{Bj}, P_B)$ are relatively close to each other. The figure verifies again equation (7b). In the most characteristic domain $0.5 \leq p_F \leq .85$ (see latter at ATP Top-200 part) , the curves are relatively flat. As it was already mentioned, the order of serves is indifferent, so B-type rule terminates the difference between consecutive games, which is too much against the traditions in tennis. That is why we reject rule B.

**Comparison with ATP Top-200 statistics**

The ATP organization on its web-site AtpWorldTour publishes statistical data regarding the period 1991- Dec, 2016, for the top 200 players, whether they are active or retired from the game. Without much details about metadata, i.e. about the way they define concepts and acquire data they release parameters $p_f^{in}$, $p_f^{Won}$, $p_s^{Won}$, $P_T^{Won}$. Their meaning, which looks more or less self-evident, can be readily found at the Notation section of the paper. The notation with indexes was introduced by us. Here small letter f and s stand for the first or second serve, unlike F or S , which stand for the player's name. The data are rounded to 1 percent accuracy by ATP. Good correlation, between the T-type theoretical model ( formula (8)) and actual empirical observations would mean the approximate equation

$$P_T^{Won} \approx P_T\left(p^{emp}\right) \quad where$$

$$p^{emp} = p_f^{in} \cdot p_f^{Won} + (1 - p_f^{in}) \cdot p_s^{Won} \qquad (21)$$

for the data belonging to each player separately.

First we selected a random number between 1 and 20, which turned out to be 7, and then selecting 10 players of ranking 7, 27, 47,...187; we obtained data and additional information according to Figure 7. The graph part of the figure shows the $P_T(p)$ curve and the $\left(p^{emp}\ , P_T^{Won}\right)$ point pairs of (21) . The distance between the curve and the points are rather small ,and, we conclude from this nice agreement, that despite all the tactical elements in tennis ( like extra effort from the server in case of a break point or like some games are used for a little rest on the side of the receiver, see (Fernandez at al. 2006) ) the assumption, that 'points are independently and equally distributed' is a useful model of tennis (Klaassen at al. 2001).

A little specific feature that we have got in Figure 7, is that all the points approximate the curve from downward, which is against the general perception of standard deviation. One of the potential explanations is, if in the ATP statistics the double fault is logged separately, rather than as a lost point behind a second serv. Indeed, introducing a further 'empirical' variable $p^{dbl_f}$ , which cannot be found at ATP web-site, and which is evidently small and typically could be in the range of $0.01 \le p^{dbl_f} \le 0.02$ . Detailed analysis, containing only elementary steps for correction of $p^{emp}$ leads to approximate equation $p^{emp} - p^{emp,corr} = p^{dbl_f} \cdot p^{emp}$ , and that would shift the $\left(p^{emp}\ , P_T^{Won}\right)$ point pairs horizontally to the left (working with $p^{emp} = 0.63$ ) by (.0063,.0126) probability unit, and, that would result in a two sided approximation. However, this assumption is based on speculation and currently effort is made to contact ATP statisticians.

### *Shaping the game and comparing existing and proposed rules*

Since the usefulness of modelling the game is justified above, now we have the right to adjust the probability of the outcome of a game to certain requirements. For this purpose , this time in a somewhat subjective way, but covering the whole range, from the top and from the bottom of the list of top 200, we selected two players, namely R. Federer with parameters ( $p_f^{in}$ , $p_f^{Won}$ , $p_s^{Won}$ , $P_T^{Won}$ ) = ( .62, .77, .57, .88) and T. Gabashvili with parameters ( $p_f^{in}$ , $p_f^{Won}$ , $p_s^{Won}$ , $P_T^{Won}$ ) = ( .57, .70, .48, .74). From (21) and (8) the corresponding $\left(p^{emp}\ , P_T(p^{emp})\right)$ points are (.694, .888) and (.605, .756) respectively as it is marked in Figure 8.

Our aim is to reduce the existing $0.756 \le P_T(p) \le 0.888$ interval, which is deemed too predictable to interval

$$0.60 \le P_T(p) \le 0.75 \qquad (22)$$

, taking into consideration on the one hand the tradition ( $P_T$ should be definitely higher than 0.5) and on the other hand the excitement of the crowd, which requires lower predictability of the outcome of the game ( $P_T$ must not be allowed high, say higher than 0.75 ). Also we select an interval, the length of which is close to that of the original interval. From (8) we get that requirement (22) is fulfilled in case $p_{trad} =: 0.537 \le p \le 0.617 := p_{exc}$ , as in Figure 8.

The way we want to obtain a lower chance of the player to win the game is to reduce his advantage as a server, by reducing the number of first serves allowed. Let x be, the number of the first x points, when in case of a first serve fault, a second serve is allowed (i.e. after x rally, a first serve fault concludes in the immediate loss of the point). The real behavior of a player is known of course, for both situations; up to x point $p = p^{emph.}$ as at (21), after that $p = p_s^{Won}$

The unknown x is to be determined in a way that the weighted average of these two base-chances $p^{emp}$ and $p_s^{Won}$ would give the boundaries of the thought for ( $p_{trad}$ , $p_{exc}$ ) interval for the lower and higher ranked players respectively. The number of rallies is subject to change, so in the calculation they are substituted by their mathematical expectations. In this way we get equations

$$\frac{x}{E_T} p^{emp} + \frac{E_T - x}{E_T} p_s^{Won} = \begin{cases} p_{trad} & \text{for Gabashvili} \\ p_{exc.} & \text{for Federer} \end{cases} \tag{23}$$

, where $E_T$ is determined by (10); $E_T(p_{trad}) = 6.70$ and $E_T(p_{exc.}) = 6.38$ .From (23) x is expressed, e.g. for the lower estimation (Gabashvili's) case

$$x = \frac{p_{trad} - p_s^{Won}}{p^{emp} - p_s^{Won}} \cdot E_T \tag{24}$$

, which gives x=3.07 for Gabashvili's parameters and x=2.42 for that of Federer. This has to be rounded to an integer, so we conclude **that only the first three service points are to be played by current rules, then a second serve chance is to be cancelled.**

We made some approximations at the calculation of (24) ( e.g. $E_T$ is determined by T-model, rather than on the basis of the new rule) and, also we made rounding of x to three, so it is necessary to check the effect of the proposed changes.

In fact, model C with formulas (17)-(20) were created after the analysis represented by (24) had been done, and in this C-type tennis evidently the empirical data are to be applied for each individual player of different ranking by substituting $p_F = p^{emph}$ and $p_S = p_s^{Won}$ . Figure 9 shows comparison between T and C-type game, i.e. between the existing and proposed rules, applied for the data of the ATP players ranked 7, 27,107,187 according to Figure 7. ( Two players from the top, because the variability of base data is higher at the top.) The 4th column proves, that indeed we remain in the $0.586 \le P_C(p_F, p_S) \le 0.749$ interval, close to that of the requirements (22). Furthermore, we can see that the chance of a break point, the average number of rallies and break points become non-negligibly higher. The new game is predicted to be somewhat longer.

In our view, the current high predictability of the game as part of the set is necessarily subject to modification in the not far future. Calculations of this paper prove that a relatively small change in the rules can help to solve the problem. The proposed rule needs, of course, practical checking in some experimental competitions. There is no need to change the setting of the height of the net, or play with balls with different properties (like heavier or less bouncy) or adjusting the hardness of the playing surface. Of course for some of the players, those with huge first serve, the modification is harmful, while for some others it could be beneficial. However, the proposal altogether has a conservative character. Probably, the proposed change creates less perturbation in tennis at other different levels (WTA, juniors, etc.).

## *Conclusion*

The Markov chain model of tennis reliably describes the outcome of a tennis game, and so it is suitable to study the effects of possible different rules for the game. Overwhelming advantage of the server, at least in top ATP tennis is a negative development primarily for the audience, but to some extent perhaps to players as well. Detailed analysis here supports, that the proposed simple and conservative change in the rules, namely, that a **second serve** ( in case of a first serve fault) **is to be allowed only at the first three points** is a remedy of the problem.

## *NOTATION*

F and S — name of the two players, F serves first, S serves second (S also means Single serve in case of a C-type game)

A, T, Bj — ( Bj(1) and Bj(2) are equivalent versions ), type of games

B and C — ( B(1) and B(2) are equivalent versions ), type of games

$p_F$ or $p$ — probability, that player F wins the point

$p_S$ — probability, that player F wins the point when player S serves

$P_A(p)$: Probability of the event, that player F wins the A-type game ( always in the subindex is the actual game type)

$P_T^{br}(p)$: Probability of the event, that a break point (at least one) does occur during a T-type game

$E_B(p_F, p_S)$: Mathematical expectation of the number of rallies (points) in a game (of B-type)

$E_T^{br}(p)$: Mathematical expectation of the number of break points during the game

Observed relative frequency of the event that:

$p_f^{in}$ first serves is in

$p_f^{Won}$ point won by the server , behind a successful first serve

$p_s^{Won}$ point won by the server , behind a second serve

$P_T^{Won}$ the game won by the server

*References*

P. K. Newton and K. Aslam. 2009. “Monte Carlo tennis: A stochastic Markov chain model" Journal of Quantitative Analysis in Sports, vol. 4, no. 3, p. 7

Madurska, Agnieszka M. 2012. “A set-by-set analysis method for predicting the outcome of professional singles tennis matches” MEng project , Imperial College London, Department of Computing

A. J. O'Malley. 2008 . “Probability formulas and statistical analysis in tennis", Journal of Quantitative Analysis in Sports, vol. 4, no. 2, p. 15

http://www.atpworldtour.com/en/stats/1st-serve/all/all/all/

W. Feller. 1971. “ An Introduction to Probability Theory and its Applications”, John Wiley & Sons, New York, Chapter 14.

G. Szigeti and V. Komornik. 2017. " Random walk with absorbing barriers in complex form", 2017 , submitted.

J.Fernandez and A. Mendez-Villanueva and B. M. Pluim. 2006. "Intensity of tennis match play" British Journal of Sports Medicine 40, pp. 387-391

F. J. G. M. Klaassen and J. R. Magnus. 2001. "Are points in tennis independent and identically distributed? Evidence from a dynamic binary panel data model", Journal of the American Statistical Association, vol. 96, no. 454, pp. 500-509

R.W.Schutz. 1970. "A mathematical model for evaluating scoring systems with specific reference to tennis" Res. Quart.for Exercise and Sport, 41, pp.552-561

T.J. Barnett and S.R. Clarke. 2002. "Using Microsoft Excel to Model a Tennis Match" Proceedings of the Sixth Australian Conference on Mathematics and Computers in Sport, Bond University Queensland, Australia , 1-3 July 2002, pp. 63-68

T. Barnett and D. Meyer and G. Pollard. 2008. "Applying match statistics to increase serving performance" Medicine and Science in Tennis, Vol.13, No 2, pp 24-27

P. Norton and S.R.Clarke. 2002. "Serving up some grand slam tennis statistics" Proceedings of the Sixth Australian Conference on Mathematics and Computers in Sport, Bond University Queensland, Australia , 1-3 July 2002, pp. 202-209

G. Pollard and K. Noble. 2002. "The characteristics of some new scoring systems in tennis" Proceedings of the Sixth Australian Conference on Mathematics and Computers in Sport, Bond University Queensland, Australia , 1-3 July 2002, pp.221-226

| | | | | | | | | | | | | | | | | | | |
|---|---|---|---|---|---|---|---|---|---|---|---|---|---|---|---|---|---|---|
| | ↓ | | ↓ | | ↓ | | ↓ | | ↓ | | ↓ | | ↓ | | ↓ | | | Identification of the game |
| F | F | F | F | F | F | F | F | F | F | F | F | F | F | F | F | .. | .. | **A type deuce game** |
| F | S | F | S | F | S | F | S | F | S | F | S | F | S | F | S | .. | .. | **Bj(1) type (alternating) deuce game** |
| F | S | S | F | F | S | S | F | F | S | S | F | F | S | S | F | .. | .. | **Bj(2) type (tie-break) deuce game** |
| | | | | | | | | | | | | | | | | | | |
| F | F | F | F | F | F | F | F | F | F | F | F | F | F | F | F | .. | .. | **T type real tennis game** |
| F | S | F | S | F | S | F | S | F | S | F | S | F | S | F | S | .. | .. | **B(1) type (alternateing) tennis game** |
| F | S | S | F | F | S | S | F | F | S | S | F | F | S | S | F | .. | .. | **B(2) type (tie-break) tennis game** |
| F | F | F | S | S | S | S | S | S | S | S | S | S | S | S | S | .. | .. | **C type (recommended) tennis game** |
| | | | ↑ | ↑ | ↑ | | ↑ | | ↑ | | ↑ | | ↑ | | ↑ | | | |
| 1 | 2 | 3 | 4 | 5 | 6 | 7 | 8 | 9 | 10 | 11 | 12 | 13 | 14 | 15 | 16 | | | number of rallies |

↓ ↑ potential end points

F S repeating unit

F Name of the first serving player

S Name of the other player

Fig. 1 Order of services in real and potential tennis games

(a)

| -2 | -1 | 0 | +1 | +2 |
|---|---|---|---|---|
| | | | | |
| 0 | 0 | 1 | 0 | 0 |
| 0 | p | 0 | q | 0 |
| $p^2$ | 0 | 2pq | 0 | $q^2$ |
| | .. | | .. | |
| .. | | .. | | .. |
| | .. | | .. | |

(b)

| -2 | -1 | 0 | +1 |
|---|---|---|---|
| | | | |
| 0 | 0 | 1 | 0 |
| 0 | p | 0 | q |
| $p^2$ | 0 | pq | 0 |
| | .. | | .. |
| .. | | .. | |
| | .. | | .. |

Fig. 2

(a) A-type Markov process, until returning to the initial (deuce) state

(b) Markov process for calculating the probability of a break-point occurrence

| -4 | -3 | -2 | -1 | 0 | +1 | +2 | +3 | +4 |
|---|---|---|---|---|---|---|---|---|
| | | | | | | | | |
| 0 | 0 | 0 | 0 | 1 | 0 | 0 | 0 | 0 |
| 0 | 0 | 0 | $p$ | 0 | $q$ | 0 | 0 | 0 |
| 0 | 0 | $p^2$ | 0 | $2pq$ | 0 | $q^2$ | 0 | 0 |
| 0 | $p^3$ | 0 | $3p^2q$ | 0 | $3pq^2$ | 0 | $q^3$ | 0 |
| $p^4$ | 0 | $4p^3q$ | 0 | $6p^2q^2$ | 0 | $4pq^3$ | 0 | $q^4$ |
| | $4p^4q$ | 0 | $10p^3q^2$ | 0 | $10p^2q^3$ | 0 | $4pq^4$ | |
| | | $10p^4q^2$ | 0 | $20p^3q^3$ | 0 | $10p^2q^4$ | | |
| | | | .. | | .. | | | |
| | | .. | | .. | | .. | | |
| | | | .. | | .. | | | |
| | | .. | | .. | | .. | | |

Fig. 3
T-type Markov process until first deuce state

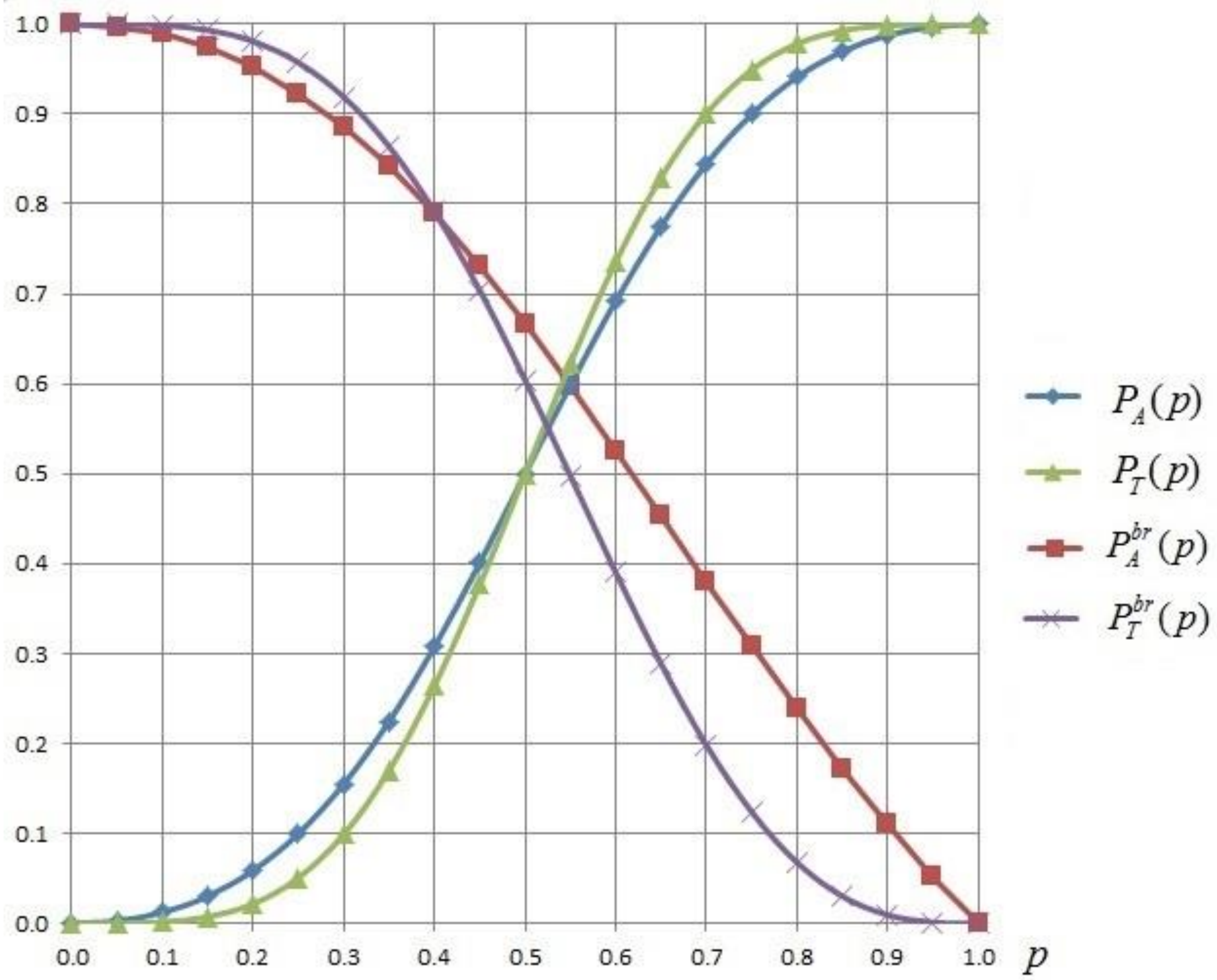

Fig. 4 Curves of base probabilities

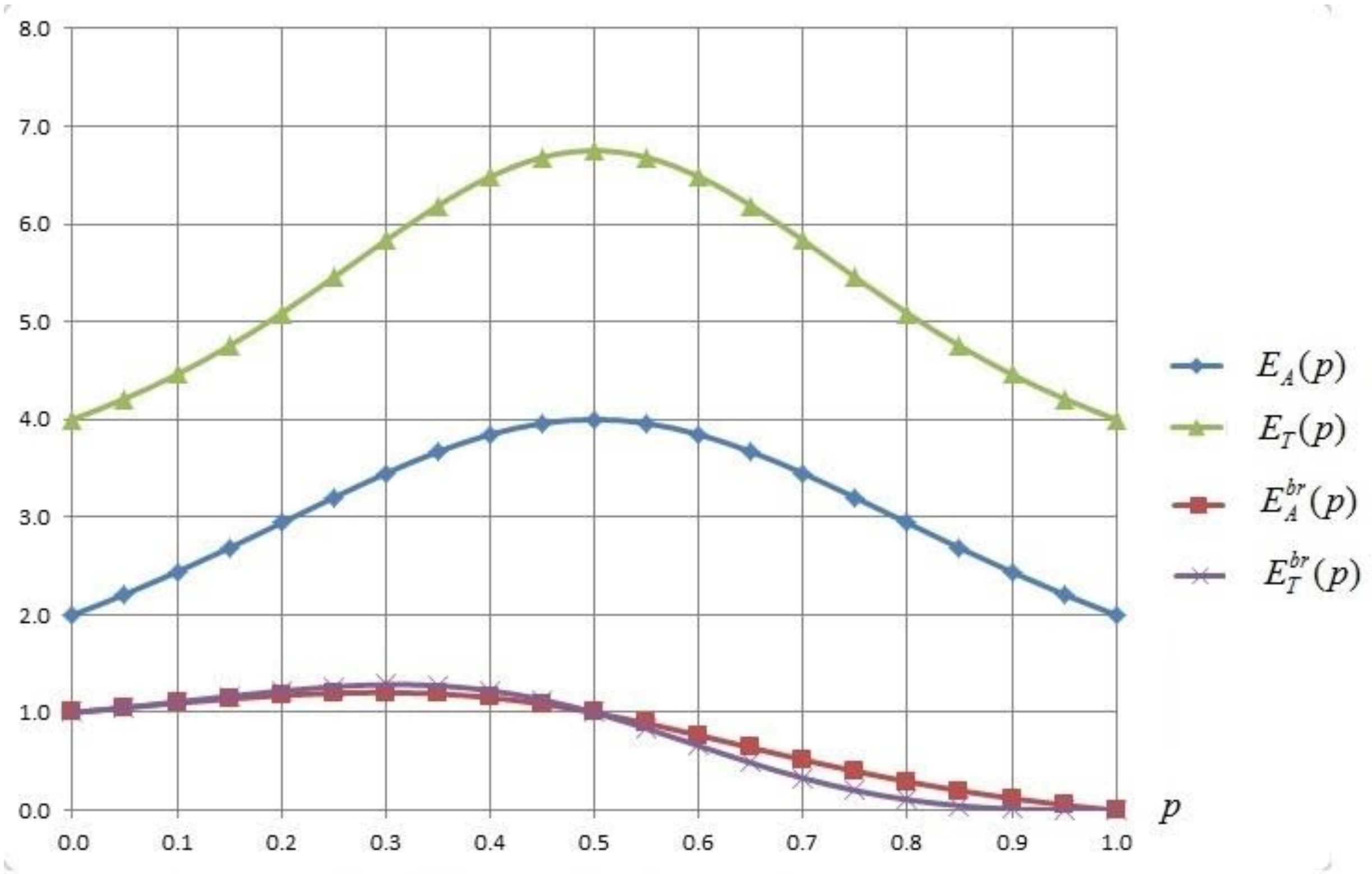

Fig. 5 Average number of rallies and break points in a game

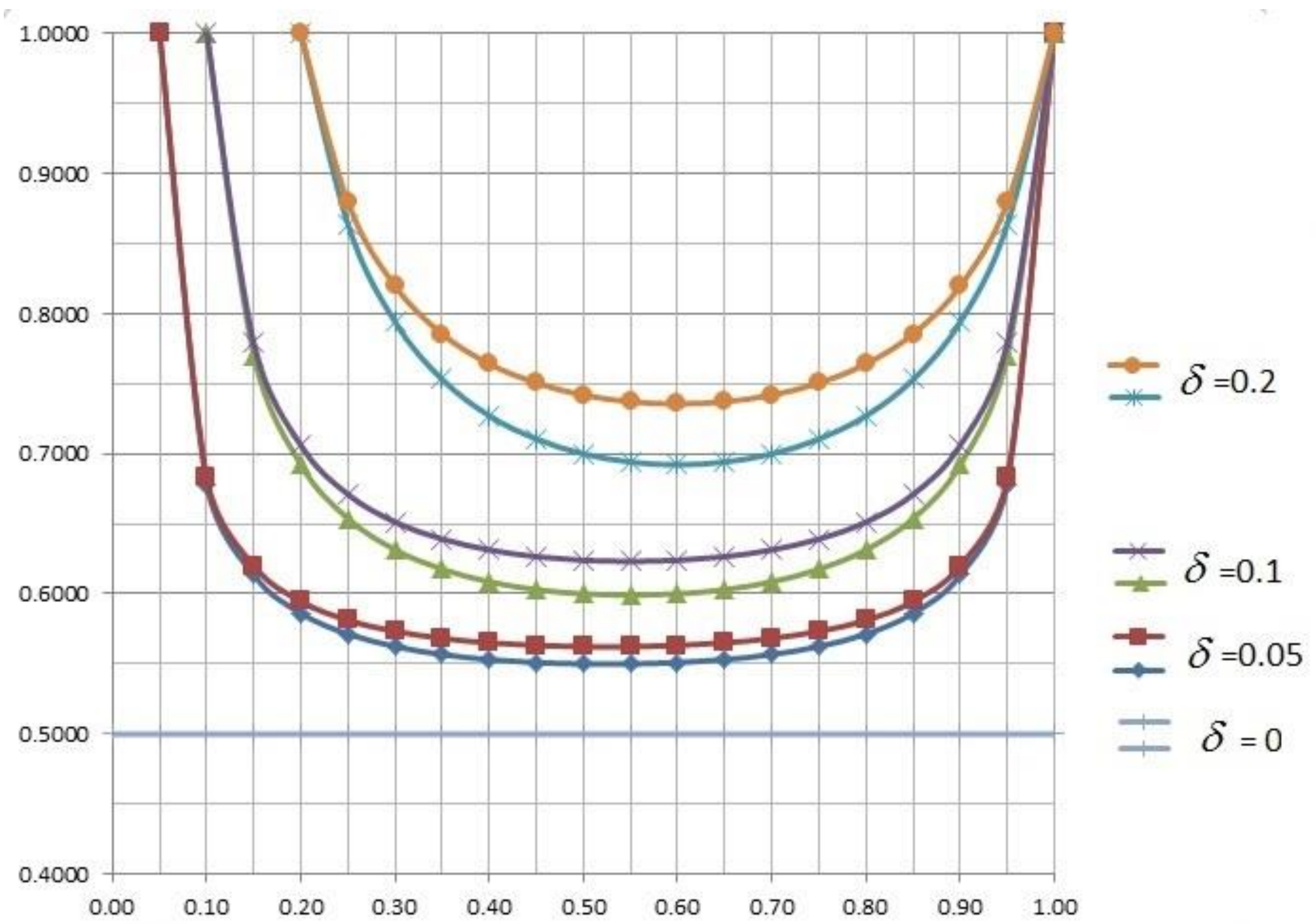

Fig. 6 Pair of curves $P_{Bj}(p_F, 1-p_F+\delta) \leq P_B(p_F, 1-p_F+\delta)$ for different small values of $\delta$

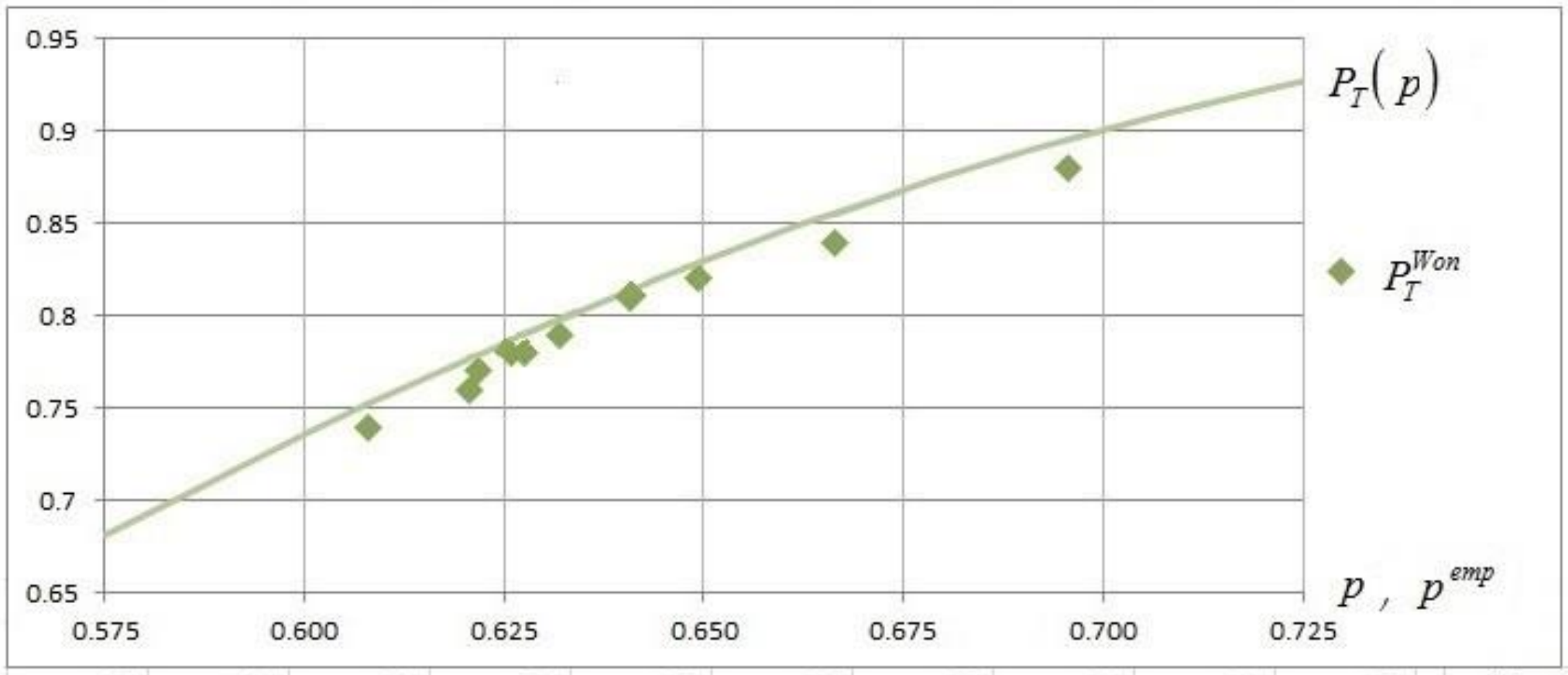

| | | | | | | | | | | | |
|---|---|---|---|---|---|---|---|---|---|---|---|
| | 274 | 463 | 280 | 202 | 558 | 629 | 227 | 228 | 268 | 575 | match played |
| | Arturs | Soderling | Stark | Shelton | Grejan | Melzer | Draper | Gofen | Lee | G.Lopez | Player's name |
| | AUS | SWE | USA | USA | FRA | A | AUS | BEL | S.KOR | SP | Nationality |
| obs. | 0.56 | 0.61 | 0.59 | 0.58 | 0.56 | 0.61 | 0.56 | 0.58 | 0.63 | 0.59 | $p_f^{in}$ |
| obs. | 0.81 | 0.76 | 0.76 | 0.75 | 0.72 | 0.70 | 0.72 | 0.71 | 0.68 | 0.69 | $p_f^{Won}$ , |
| obs. | 0.55 | 0.52 | 0.49 | 0.49 | 0.52 | 0.51 | 0.51 | 0.50 | 0.52 | 0.49 | $p_s^{Won}$ , |
| obs. | **0.88** | **0.84** | **0.82** | **0.81** | **0.79** | **0.78** | **0.78** | **0.77** | **0.76** | **0.74** | $P_T^{Won}$ |
| calc. | 0.696 | 0.666 | 0.649 | 0.641 | 0.632 | 0.626 | 0.628 | 0.622 | 0.621 | 0.608 | $p^{emp}$ |
| | 7 | 27 | 47 | 67 | 87 | 107 | 127 | 147 | 167 | 187 | Ranking by $P_T^{Won}$ |

obs.= observed by ATP statistician  calc.= calculated from obs. data

Fig. 7 Comparison with empirical data

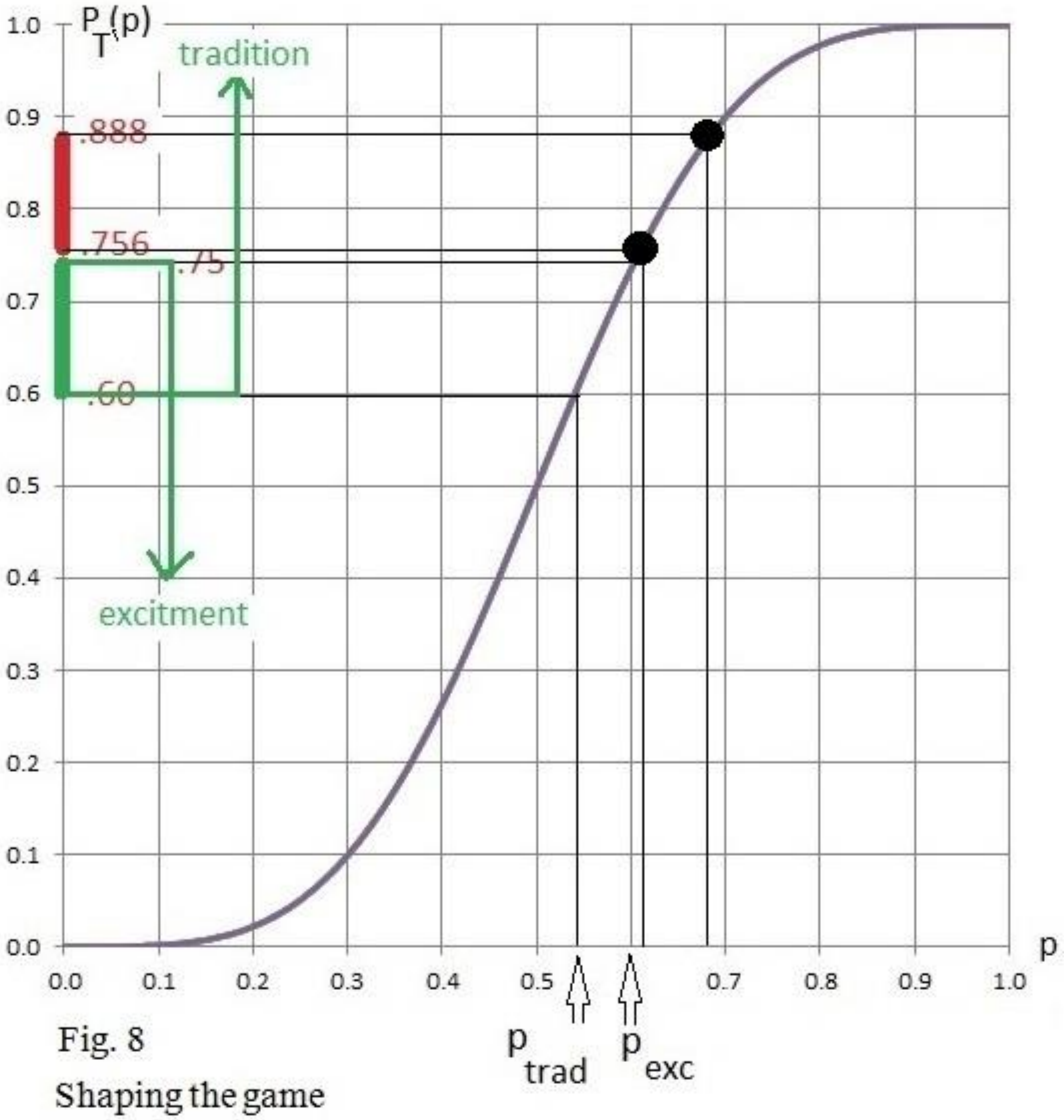

Fig. 8

Shaping the game

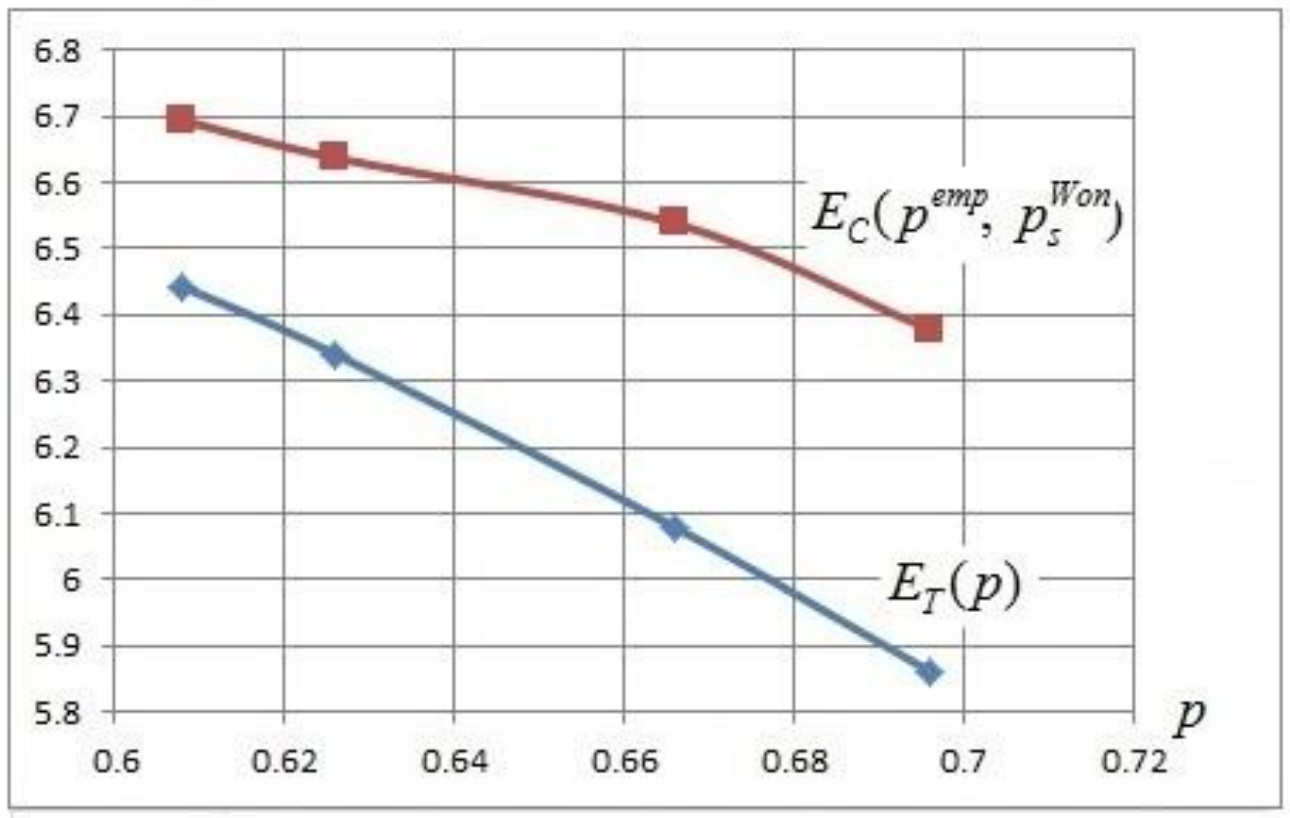

| | $p = p^{emp}$ | $p_s^{Won}$ | $P_T(p)$ | $P_C(p, p_s^{Won})$ | $P_T^{br}(p)$ | $P_C^{br}(p, p_s^{Won})$ | $E_T(p)$ | $E_C(p, p_s^{Won})$ | $E_T^{br}(p)$ | $E_C^{br}(p, p_s^{Won})$ |
|---|---|---|---|---|---|---|---|---|---|---|
| 7 | 0.696 | 0.55 | 0.896 | 0.749 | 0.205 | 0.345 | 5.861 | 6.378 | 0.918 | 1.410 |
| 27 | 0.666 | 0.52 | 0.855 | 0.683 | 0.258 | 0.410 | 6.079 | 6.538 | 1.027 | 1.512 |
| 107 | 0.626 | 0.51 | 0.787 | 0.633 | 0.336 | 0.463 | 6.340 | 6.637 | 1.173 | 1.541 |
| 187 | 0.608 | 0.49 | 0.752 | 0.586 | 0.373 | 0.505 | 6.443 | 6.694 | 1.237 | 1.600 |

Fig.9 Comparison of existing and proposed games